\documentclass[twocolumn,aps,amsmath,amssymb,prb]{revtex4} 
\usepackage{graphicx}

\renewcommand{\L}{{\cal L}}

\newcommand{\G}{{\cal G}}

\newcommand{\B}{\mbox{\tiny B}}
\newcommand{\Sec}[1]{Sec.\,\ref{#1}}
\newcommand{\ti}{\tilde}

\newcommand{\be}{\begin{equation}}
\newcommand{\ee}{\end{equation}}
\newcommand{\bea}{\begin{eqnarray}}
\newcommand{\eea}{\end{eqnarray}}
\newcommand{\bsube}{\begin{subequations}}
\newcommand{\esube}{\end{subequations}}
\newcommand{\Eq}[1]{Eq.~(\ref{#1})}
\newcommand{\Eqs}[1]{Eqs.~(\ref{#1})}
\newcommand{\Fig}[1]{Fig.\,\ref{#1}}

\newcommand{\la}{\langle}
\newcommand{\ra}{\rangle}

\newcommand{\kT}{\mbox{$k_{\rm B}T$}}

\begin{document}

\title{Electron transfer theory revisit: Quantum solvation effect}
\author{Ping Han,$^{a,c)}$ Rui-Xue Xu,$^{b,c)\ast}$
        Ping Cui,$^{c)}$
          Yan Mo,$^{c)}$
        Guozhong He,$^{a)}$
 }
\author{YiJing Yan$^{a,b,c)}$}
\email{rxxu@ustc.edu.cn; yyan@ust.hk}
\affiliation{$^{a)}$State Key Laboratory of Molecular Reaction Dynamics,
  Dalian Institute of Chemical Physics,
  Chinese Academy of Sciences, Dalian 116023, China \\
$^{b)}$Hefei National Laboratory for Physical Sciences
  at the Microscale, University of Science  and Technology of China,
  Hefei 230026, China    \\
$^{c)}$Department of Chemistry,
    Hong Kong University of Science and Technology, Kowloon, Hong Kong}

\date{Accepted 25 May 2006,
{\it J.\ Theo.\ \& Comput.\ Chem.}; manu\#: jtcc06052a}

\begin{abstract}
 The effect of solvation on ET rate processes is 
 investigated on the basis of the exact theory constructed
 in J.\ Phys.\ Chem.\  {\bf B 110}, xxx (2006).
 The nature of solvation is studied in a close
 relation with the mechanism of ET processes.
 The resulting  Kramers' turnover and Marcus' inversion 
 characteristics are analyzed accordingly.   
 The classical picture of solvation is found to be invalid
 when the solvent longitudinal relaxation time is short
 compared with the inverse temperature.
\end{abstract}

\maketitle

\section{Introduction}
\label{thintro}

   Chemical reaction in condensed phases is
 intimately related to the Brownian motion in solution.
 Einstein's paper on Brownian motion\cite{Ein05549}
showed the first time the fluctuation-dissipation relation (FDR).
 The fluctuations of surrounding molecules 
 are responsible for both agitation and
 friction on the Brownian particle.
 These stochastic events of energy exchange between system
 and bath lead eventually to thermal equilibrium.
 Brownian motion is characterized by the stochastic force.
 The FDR leads to a Fokker-Planck equation, which is the classical
 reduced equation of motion that governs the Brownian 
 motion at an ensemble average level.
 This approach has been exploited by Kramers
 in his construction of isomerization
 reaction rate theory.\cite{Kra40284}
  The resulting rate is shown to have a maximum
 in an intermediate viscosity region. This celebrated
 Kramers' turnover behavior clearly demonstrates
  the dual role of solvent on 
 reaction rate.\cite{Kra40284,Han90251}
 
  Electron transfer (ET) is the simplest 
 but a pivotally important chemical reaction system. It constitutes 
 another class of systems whose dependence on solvent 
 environment has been extensively studied since the pioneering
 work by Marcus in 1950s.%
\cite{Mar56966,Mar64155,Mar85265,Zus80295,Zus8329,Hyn85573,%
Gar854491,Fra85337,Wol871957,Spa873938,Spa883263,Spa884300,%
Yan884842,Yan896991,Yan979361,Tan973485,Bix9935}
 However, it is often treated in different way from 
 the traditional chemical reaction involving bond breaking and/or
 formation. In the latter case, either the equation of motion for 
 a particle over the barrier or the flux-flux 
 correlation function approach on the basis of 
 the transition-state theory is used.%
\cite{Han90251,Mil981,Fre02,Pol05026116} 
 The standard treatment in the ET research field 
 is rather a type of transfer coupling 
 correlation function formalism, based on the
 assumption that the nonadiabatic coupling matrix element 
 $V$ is not strong.%
\cite{Mar56966,Mar64155,Mar85265,Zus80295,Zus8329,Hyn85573,%
Gar854491,Fra85337,Wol871957,Spa873938,Spa883263,Spa884300,%
Yan884842,Yan896991,Yan979361,Tan973485,Bix9935}

 Depicted in \Fig{fig1} is the schematics
 of an elementary donor-acceptor ET system. 
 Here, $E^{\circ}$ denotes the reaction endothermicity;
 $V_a$ ($V_b$) represents the potential surface of the solvent environment for
 the electron in the donor (acceptor) state;
 $U\equiv V_b-V_a-E^{\circ}$ is the solvation coordinate;
 while $\lambda \equiv \la U \ra$ is the solvation energy, with
 $\la \cdots \ra$ denoting the initial bath ensemble average.
 At the crossing ($U+E^{\circ}=0$) point,
 $V_a=V_b=(E^{\circ}+\lambda)^2/(4\lambda)$ that amounts to
 the ET reaction barrier height. The celebrated Marcus' rate theory 
 reads\cite{Mar56966,Mar64155,Mar85265}
 \be \label{mark0}
  k = \frac{V^2/\hbar}{\sqrt{\lambda\kT/\pi}}
      \exp\left[-\frac{(E^{\circ}+\lambda)^2}{4\lambda\kT}\right].
 \ee
 It is a classical Franck-Condon theory, assuming that
 the solvent relaxation is much slow 
 compared with the electronic transition.
 Exploited in \Eq{mark0} is also the classical FDR:
  $\la U^2 \ra - \la U\ra^2 = 2\kT\la U \ra$.
 Quantum extension of Marcus' theory
 has been formulated in the weak transfer coupling 
 regime.\cite{Tan973485,Bix9935} 
 The dynamic solvation effect is introduced by 
 the solvation correlation function,
\be \label{Ct}
   C(t) = \la U(t)U(0) \ra - \la U \ra^2.
\ee
 Nonperturbative rates have also been 
 formulated on the basis of
 fourth-order transfer correlation functions,
 followed by certain resummation schemes 
 to partially account for the nonperturbative
 transfer coupling effects.\cite{Hyn85573,%
Gar854491,Fra85337,Wol871957,Spa873938,Spa883263,%
Spa884300,Yan884842,Yan896991,Yan979361}
The resulting rates, despite of the resummation approximation
involved, do recover the celebrated
Kramers' turnover behavior.

 The main purpose of this work is to elucidate some
 distinct solvation effects on the ET rate processes.
 The quantum nature of solvation arises from the fact
 that the solvation coordinate is an operator 
 and its correlation function must be complex.
 The elementary ET system in Debye solvents will be studied
 on the basis of the exact and analytical rate 
 theory developed in Ref.\ \onlinecite{Han06jpcB}, 
 which will be referred as Paper I hereafter.
 Section \ref{ththeo} summarizes the theoretical results of Paper I.
 This is a reduced quantum equation of motion based formalism;
 i.e., the quantum version of Kramers' Fokker-Planck equation approach.
  The key quantity now
 is the reduced density matrix, 
 $\rho(t)=\text{tr}_\text{\B}\rho_{\rm T}$,
 defined as the trace of the total system-bath density matrix
 over the bath subspace.  
 Numerical results will be presented and discussed 
 in \Sec{thnum}. Finally, \Sec{thsum} concludes this paper. 

\section{An exact and analytical theory}
\label{ththeo}

 This section summarizes the exact and analytical rate 
 theory, developed in Paper I, for the ET in Debye 
 solvents at finite temperatures.
 Let us start with the following form
 of the reduced Liouville equation,
 \be \label{cop}
   \dot\rho(t)=-\frac{i}{\hbar}[H,\rho(t)] - 
  \int_{0}^t \!\! d\tau\, \hat\Pi(t,\tau)\rho(\tau).
 \ee
 For the present ET system (\Fig{fig1}), 
 the reduced system Hamiltonian reads
 \be \label{Hsys}
    H = (E^{\circ}+\lambda)|b\ra\la b| + V(|a\ra\la b|+|b\ra\la a|).
 \ee
 This is a time-independent system, for which the 
 dissipation memory kernel $\hat\Pi(t,\tau) = \hat\Pi(t-\tau)$.
 As a result, \Eq{cop} can be resolved in its Laplace domain as 
 \be \label{rhos}
   s\ti\rho(s) -\rho(0) = -i\L\ti\rho(s) - \Pi(s)\ti\rho(s).
 \ee
 Here $\L\equiv \hbar^{-1}[H,\bullet]$ is the reduced system
 Liouvillian.  

   A simplification arises for the ET system in Debye solvents at 
 finite temperature. The solvation correlation 
 function assumes now (for $t>0$)
 \be \label{debyeC}
   C(t) = \lambda(2\kT-i\hbar\gamma) e^{-\gamma t}
   \equiv \hbar^2 \eta e^{-t/\tau_{\rm L}}.
 \ee
 Here, $\tau_{\rm L}\equiv \gamma^{-1}$ denotes
 the longitudinal relaxation time of the Debye solvent.
 In this case, \Eq{rhos} can be formulated exactly in terms of 
 a continued fraction Green's function formalism.
 Let $\Pi(s)\equiv \Pi^{(0)}(s)$   
 and
 \be \label{Gs}
    \G^{(n)}(s) \equiv  \frac{1}{s+i\L+\Pi^{(n)}(s)}; \ \ n\geq 0.
 \ee
 The continued fraction hierarchy is now the relation
 between $\Pi^{(n)}(s)$ and $\G^{(n+1)}(s+\gamma)$;
 cf. the eq (16) of Paper I.
  For the elementary ET system subject to the Debye longitudinal 
 relaxation, it is found that $\Pi^{(n)}$, which is Hermite
  satisfying $\Pi^{(n)}_{jj',kk'}=\Pi^{(n)\,\ast}_{j'j,k'k}$, 
 has only three nonzero tensor elements
 together with their complex conjugates. 
 Denote the three nonzero tensor elements of $\Pi^{(n)}$ as
\be\label{xyz}
   x \equiv  \Pi_{ba,ba}, \
   y \equiv  \Pi_{ba,ab}, \
   z \equiv  \Pi_{ba,bb}.
\ee
 Implied here, and also whenever applicable hereafter 
 [cf.\ \Eqs{finalX}--(\ref{beta})], 
 are the superscript index $(n)$ and argument $s$,
 if they are the same in the both sides of 
 the individual equation; otherwise they will be specified.
 The continued fraction hierarchy that 
 relates $\Pi^{(n)}(s)$ with $\G^{(n+1)}(s+\gamma)$ can now
 be expressed in terms of
 \bsube \label{finaldebye}
\bea
   x^{(n)}(s) &=&  \eta(n+1) X^{(n+1)}(s+\gamma),
    \ \  \ \
\label{finalx} \\
   y^{(n)}(s) &=& - \eta^{\ast} (n+1) Y^{(n+1)}(s+\gamma),
\label{finaly} \\
   z^{(n)}(s) &=& (\eta-\eta^{\ast})(n+1) Z^{(n+1)}(s+\gamma).
\label{finalz}
 \eea \esube
 Here, $\{X,Y,Z\}^{(n)}(s)$ are the counterpart tensor elements of 
 the Green's function $\G^{(n)}(s)$.
 Their relations to the nonzero elements $\{x,y,z\}^{(n)}(s)$ of
 $\Pi^{(n)}(s)$ via \Eq{Gs} can be evaluated analytically 
 on the basis of the Dyson equation technique. 
 The final results are
 \bsube \label{finalXYZ}
  \bea \label{finalX}
 X\!\!&\equiv&\!\!\G_{ba,ba} =
 \frac{ \alpha^{\ast}+\beta^{\ast} }
             {|\alpha + \beta|^2 - |\beta - y|^2},
 \\ \label{finalY}
 Y\!\!&\equiv&\!\!\G_{ba,ab}
 = \frac{\beta-y}{|\alpha + \beta|^2 - |\beta - y|^2},
\\ \label{finalZ}
 Z\!\!&\equiv&\!\!\G_{ba,bb}
  = -\frac{1}{s}
   \bigl[(z\!-\!iV/\hbar) X + (z^{\ast}\!+\!iV/\hbar) Y\bigr],
\eea
 \esube
with
\bsube \label{albeta}
\bea
 \alpha &\equiv&  s+(i/\hbar)(E^{\circ}+\lambda)+x,
\label{alpha} \\
  \beta &\equiv& s^{-1}(V/\hbar)^2[2+i\hbar z/V].
\label{beta}
\eea
\esube
 
   The kinetics rate equation can be readily obtained
 via \Eq{rhos} by eliminating the off-diagonal
 reduced density matrix elements. 
 It leads to a
 linear algebraic equation in the Laplace domain 
 that corresponds to the generalized rate equation with
 memory rate kernels in time domain.
 The resulting ET rate resolution
 reads as [the eq (36a) of Paper I]
 \be \label{Ks}
  k(s) = \frac{2|V|^2}{\hbar^2 }{\rm Re}
    \frac{\alpha(s) + y(s)}{|\alpha(s)|^2 - |y(s)|^2}.
 \ee
 The rate constant $k\equiv k(0)$ 
 that will be numerically studied in the next section
 amounts to the time integral of the memory rate kernel
 in the aforementioned generalized rate equation.
 
   As analyzed in Paper I,
 the infinity inverse recursive formalism [\Eqs{finaldebye}
 and (\ref{finalXYZ})] can be
 truncated by setting $\{x,y,z\}^{(N)}=0$ at 
 a sufficiently large anchoring $N$.
 The resulting $\{x,y,z\}^{(0)}\equiv \{x,y,z\}$ that are required
 by \Eq{Ks} are exact up to the $(2N)^{\rm th}$-order
 in the system-bath coupling. The convergence is guaranteed also via
 the mathematical continued fraction structure involved.
 Apparently, if the rates are needed at a specified $s'$,
 one shall start with $\{x,y,z\}^{(N)}_{s=s'+N\gamma}=0$.
 The backward-recursion relations, \Eqs{finaldebye} with \Eqs{finalXYZ},
 will then lead to $\{x,y,z\}^{(N-1)}_{s=s'+(N-1)\gamma}$, and so on,
 until $\{x,y,z\}^{(0)}_{s=s'}\equiv \{x,y,z\}_{s=s'}$
 are reached for evaluating the required $k(s')$ [\Eq{Ks}]. 
  The above reduced dynamics-based ET rate formalism is exact
 for the Debye solvents in finite temperatures.
 However, the FDR, which relates the
 real and imaginary parts of the solvation correlation
 function, is adopted in \Eq{debyeC} in a semiclassical manner.
 As a consequence the reduced density matrix and 
 rates may become negative if the temperature is too low.

\section{Quantum solvation effects:  Numerical results} 
\label{thnum}
   We are now in the position to elucidate 
 some distinct solvation effects 
 on the ET reaction rate $k\equiv k(s=0)$ [cf.\ \Eq{Ks}].
 Numerical results will be presented 
 in relation to the celebrated Kramers' turnover and 
 Marcus' inversion behaviors, exemplified with
 the ET reaction systems of $V=1$ kJ/mol 
 and $\lambda=3$ kJ/mol at $T=$ 298\,K. 

  It is noticed that the solvation longitudinal relaxation 
 time $\tau_{\rm L}$ is considered proportional to the solvent 
 viscosity.\cite{Yan884842,Yan896991}
 The Kramers' turnover characteristics can therefore be demonstrated 
 in terms of the rate $k$ as a function of the scaled
 solvent relaxation time $\tau_{\rm L}/\tau_{\rm ther}$.
 Here, $\tau_{\rm ther} \equiv \hbar/(\kT)$ denotes the thermal time,
 which at the room temperature is about 26\,fs.
   In the Debye solvent model of \Eq{debyeC},
 the quantum nature of solvation enters via 
 the semiclassical FDR that relates the real and 
 imaginary parts of the correlation
 function. In contrast, the classical solvation is
 characterized by the real part only. 
  As $\text{Im}\,\eta/\text{Re}\,\eta = -0.5 \tau_{\rm ther}/\tau_{\rm L}$,
 it is anticipated that the quantum nature of solvation
 can only be prominent in the 
 low viscosity ($\tau_{\rm L}<\tau_{\rm ther}$) regime.
 It is also consistent with
 the physical picture that the high viscosity
 (or slow motion) implies a large effective mass
 and thus leads to the classical solvation limit. 
  
  Figure \ref{fig2} depicts the rates $k$ as functions of 
  $\tau_{\rm L}/\tau_{\rm ther}$
 for two typical systems, being of the
 endothermicity values of
 $E^{\circ}=-\lambda$ and $E^{\circ}=0$, respectively.
 Observed in the high viscosity ($\tau_{\rm L}/\tau_{\rm ther}>1$) 
  regime for each of the systems is
 the celebrated Kramers' fall-off behavior.\cite{Kra40284,Han90251}
 This is the well established classical solvation picture 
 of the diffusion limit: the higher the solvent
 viscosity is, the more backscattering (or barrier-recrossing)
 events will be. The fact that $k_{E^{\circ}=-\lambda} > k_{E^{\circ}=0}$
 observed in the high viscosity regime is also 
 anticipated from the classical solvation picture 
 [cf.\ \Fig{fig1} or \Eq{mark0}]:
 That $E^{\circ}=-\lambda$ represents 
 a classical barrierless system 
 where the celebrated Marcus' inversion takes place. 

 In the low viscosity ($\tau_{\rm L}/\tau_{\rm ther}<1$) regime
 the  classical picture of solvation is however invalid.
  The observed rate in the symmetric ($E^{\circ}=0$) system, 
 is apparently tunneling dominated due to Fermi resonance.
 The most striking observation is that
 the so called barrierless ($E^{\circ}=-\lambda$) system 
 exhibits now clearly the 
 Kramers' viscosity-assisted barrier-crossing 
 characteristics in the
 low viscosity regime. This suggests that 
 there is an effective barrier for 
 the ET system with the classical 
 barrierless value of $E^{\circ}+\lambda=0$;
 this effective barrier is viscosity dependent and
 vanishes as $\tau_{\rm L}$ increases.

  Now turn to the Marcus' inversion characteristics
 for the rate $k$  as a function of reaction endothermicity $E^{\circ}$. 
 Depicted in \Fig{fig3} are the resulting inversion curves, with
 $\tau_{\rm L}/\tau_{\rm ther}=0.1, 1$, and 10
 to represent the low (solid-curve), intermediate (dot-curve), 
 and high (dash-curve) viscosity regimes,
 respectively.
  In the classical solvation picture the inversion
 occurs at $E^{\circ}=-\lambda$, as it represents
 a classical barrierless system. This picture
 is only valid in the high viscosity regime;
 see the dashed curve with $\tau_{\rm L}/\tau_{\rm ther}=10$. 

   In the low viscosity regime, 
 according to the analysis presented for \Fig{fig2},
 there is always a nonzero barrier for the ET reaction,
 covering over the entire range of $E^{\circ}$
 including the value of $E^{\circ}=-\lambda$.
 This explains the inversion behavior of
 the solid curve in \Fig{fig3} that is peaked only
 at the resonant position of $E^{\circ}=0$.
  As the viscosity increases, the inversion region
 smoothly shifts from the resonant peak position
 $E^{\circ}=0$ to the classical barrierless position
 of $E^{\circ}=-\lambda$.

   To explain the asymmetric property of the 
 inversion behavior observed in \Fig{fig3}, 
 let us recall that $k(-E^{\circ})$ amounts to the backward reaction
 rate and $k(E^{\circ})<k(-E^{\circ})$ for an endothermic
 ($E^{\circ}>0$) reaction. This leads immediately
 to the asymmetric property of the solid curve in \Fig{fig3},
 in which the blue (endothermic) wing falls off faster
 than its red (exothermic) wing.
 The degree of asymmetry decreases as the viscosity increases.
 Only in the high viscosity regime does the inversion curve
 behave classically, which is symmetric (but may not be parabolic unless
 the transfer coupling $V$ is small) 
 around its classical inversion position of $E^{\circ}=-\lambda$.
 
\section{Summary}
\label{thsum}
   In summary, we have investigated in detail the 
 effect of solvation on ET rate processes. The
 nature of solvation is studied in a close 
 relation with the mechanism
 of ET processes in terms of Kramers' turnover and
 Marcus' inversion characteristics.
  The classical picture of solvation is found to be invalid
 in the low viscosity regime, which can be well measured 
 by the scaled longitudinal relaxation time of
 $\tau_{\rm L}/\tau_{\rm ther}$, where
 $\tau_{\rm ther}=\hbar/(\kT)$ is the thermal time.
  The present study is carried out on the basis 
 of the exact rate theory for the simplest ET system
 with a single solvation relaxation time scale. 
 Nevertheless, the basic results obtained here are 
 expected to be valid to a general ET system
 in a realistic solvent environment of 
 multiple relaxation time scales.

\begin{acknowledgments}
 Support from the RGC Hong Kong and the NNSF of
 China (No.\ 50121202, No.\ 20403016 and No.\ 20533060)
 and Ministry of Education of China (no.\ NCET-05-0546)
 is acknowledged.
\end{acknowledgments}


\clearpage

\begin{figure}
\caption{
 Schematics of solvent environmental  potentials $V_a$ and $V_b$
 for the ET system in the donor and acceptor states,
 respectively, as the functions of the solvation coordinate
 $U\equiv V_b-V_a-E^{\circ}$, with
  $E^{\circ}$ being the ET endothermicity
 and $\lambda = \la U\ra$ the solvation energy.
 The classical barrierless system is that of $E^{\circ}+\lambda=0$.
}
\label{fig1}
\end{figure}

 \begin{figure}
 \caption{
    ET rates ($k$) as functions of scaled solvent 
 longitudinal relaxation time ($\tau_{\rm L}/\tau_{\rm ther}$)
 for the symmetric ($E^{\circ}=0$) and the 
 classical barrierless ($E^{\circ}+\lambda=0$)
 systems at $T$=298 K, with $\lambda=3$ kJ/mol and $V=1$ kJ/mol.
 The thermal time $\tau_{\rm ther}\equiv\hbar/(\kT)= 10^{-1.6}$ps 
 at the room temperature. 
}
 \label{fig2}
 \end{figure}

\begin{figure}
 \caption{
  ET rates ($k$) as functions of 
 reaction endothermicity ($E^{\circ}$). 
  Three values of the relative relaxation time,
  $\tau_{\rm L}/\tau_{\rm ther}=0.1, 1, 10$ are used
  to represent low, intermediate, and high viscosity regimes.
}
 \label{fig3}
 \end{figure}

\begin{center}
\centerline{\includegraphics[width= 0.85\columnwidth,angle=-90]
 {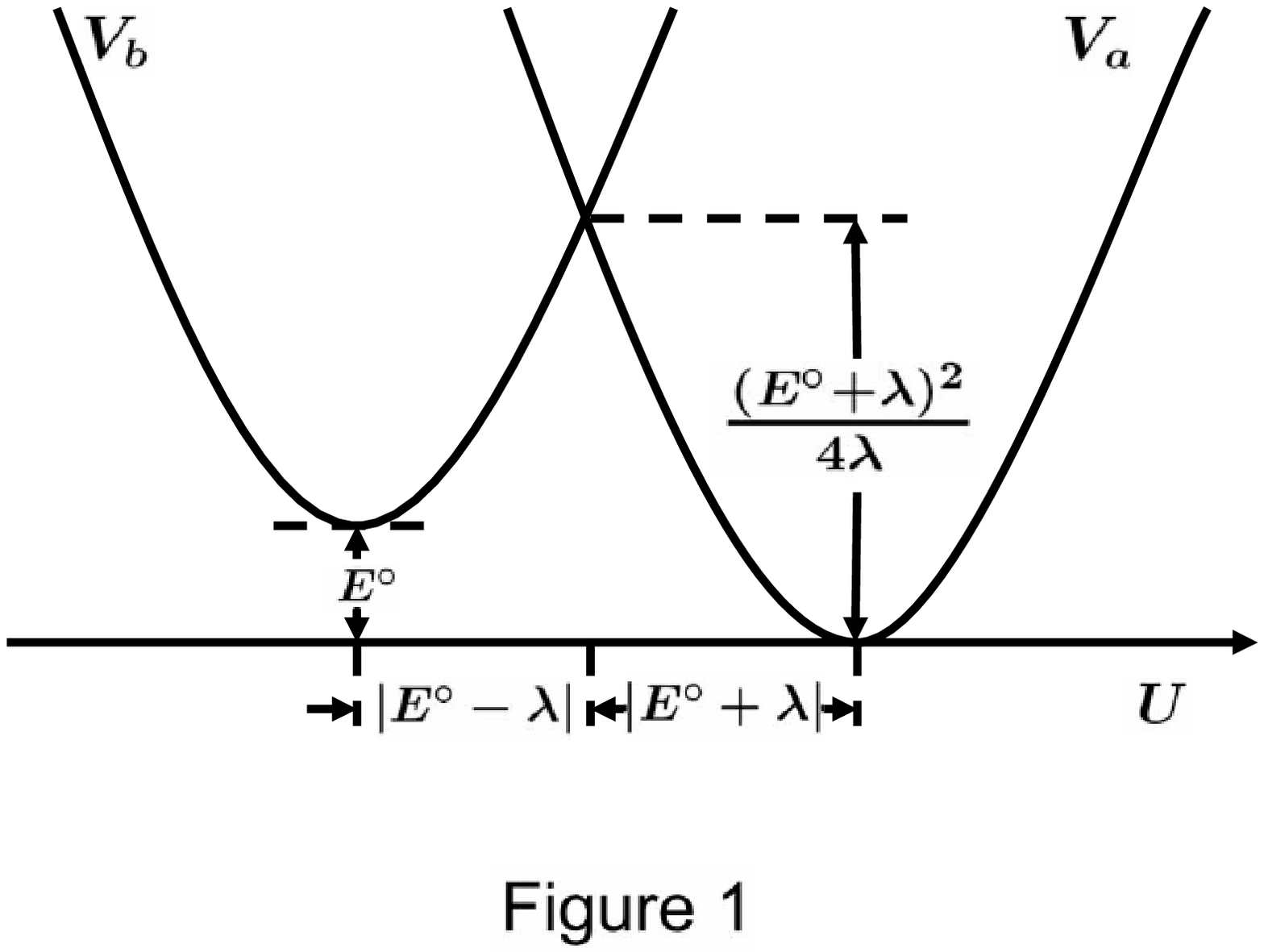}}
\centerline{\includegraphics[width= 0.85\columnwidth,angle=-90]
 {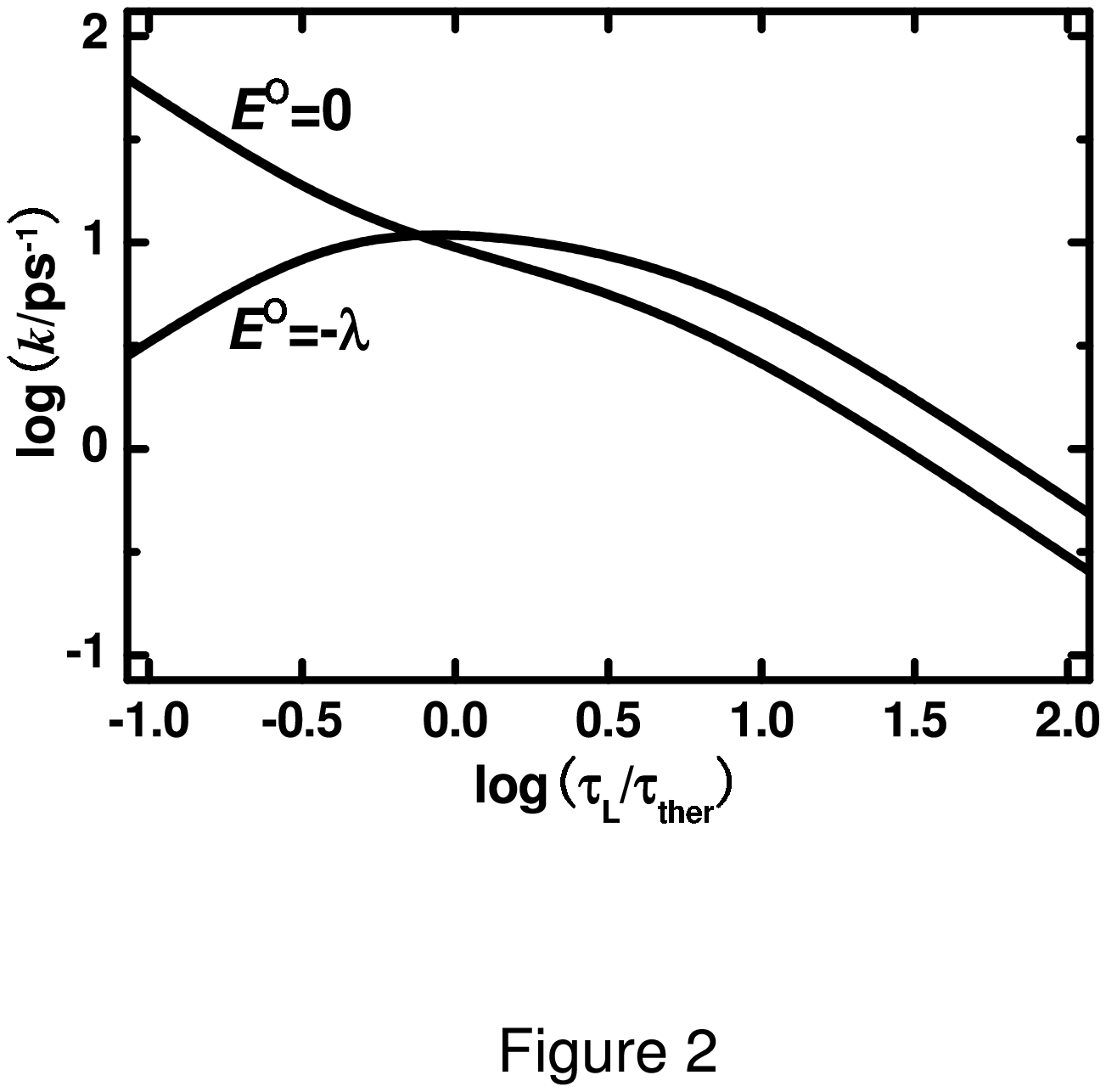}}
\centerline{\includegraphics[width= 0.85\columnwidth,angle=-90]
 {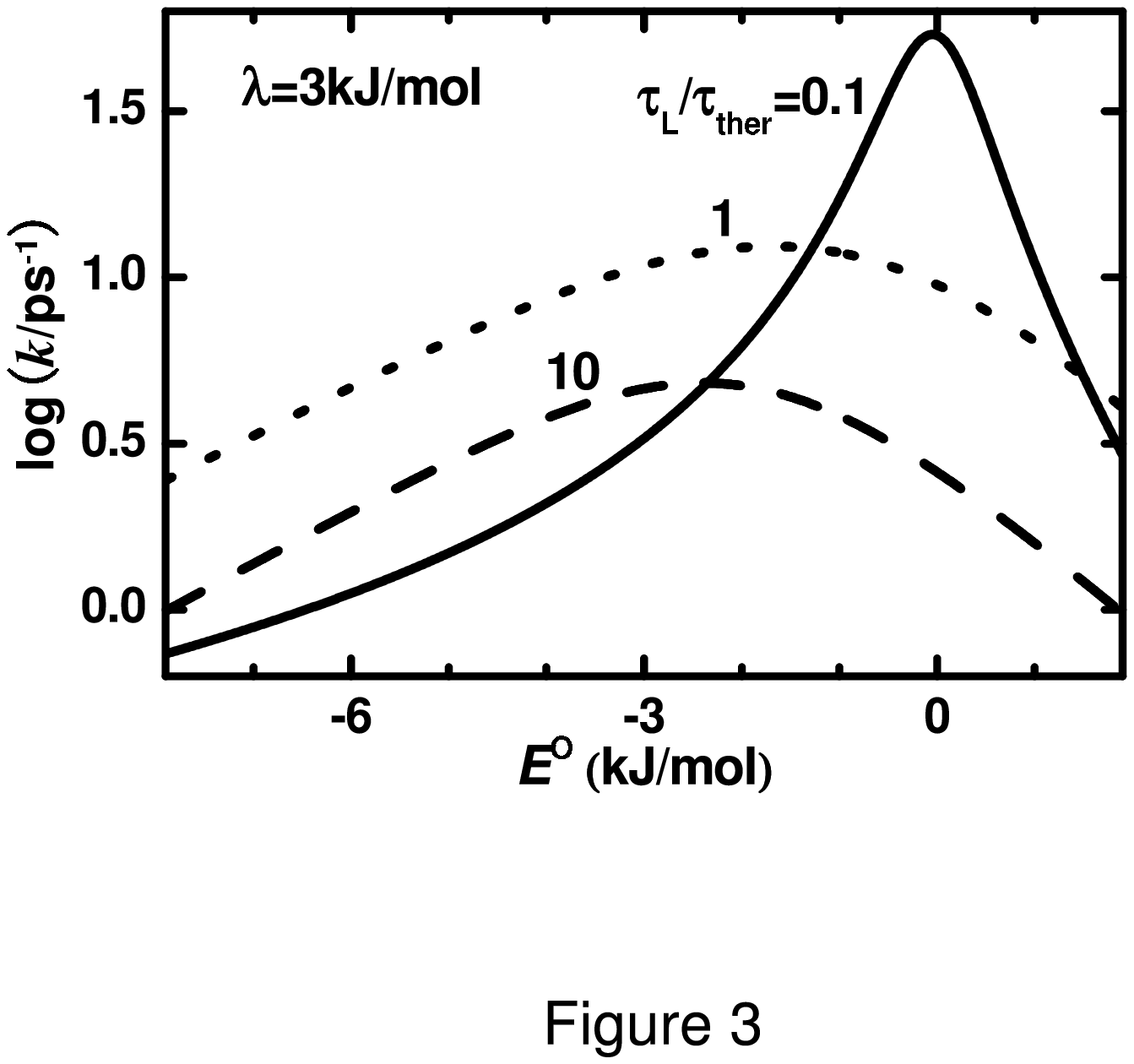}}
\end{center}

\end{document}